\documentclass{article}
\usepackage{spconf,amsmath,graphicx}

\usepackage{xcolor}
\usepackage{url}
\definecolor{darkgreen}{RGB}{1,80,22}
\usepackage{enumitem}
\usepackage{hyperref}
\usepackage{bm}
\setlist{nosep, leftmargin=14pt}
\usepackage{booktabs}
\usepackage{mwe} % to get dummy images
\usepackage{pifont}% http://ctan.org/pkg/pifont
\title{Diffusion Models with Ensembled Structure-based Anomaly Scoring for Unsupervised Anomaly Detection}
\name{
\begin{tabular}{lllll}
Finn Behrendt$^{\star }$  &
Debayan Bhattacharya$^{\star }$  &
Lennart Maack$^{\star }$  &
Julia Krüger$^{\dagger}$ & \\
Roland Opfer$^{\dagger}$&
Robin Mieling$^{\star}$ &
Alexander Schlaefer$^{\star}$
\end{tabular}}
\address{$^{\star}$ Institute of Medical Technology and Intelligent Systems,\\ Hamburg University of Technology,  Hamburg, Germany \\  $^{\dagger}$Jung Diagnostics GmbH, Hamburg, Germany}

\begin{document}
%\ninept
%
\maketitle
\begin{abstract}
Supervised deep learning techniques show promise in medical image analysis. However, they require comprehensive annotated data sets, which poses challenges, particularly for rare diseases. Consequently, unsupervised anomaly detection (UAD) emerges as a viable alternative for pathology segmentation, as only healthy data is required for training. However, recent UAD anomaly scoring functions often focus on intensity only and neglect structural differences, which impedes the segmentation performance. This work investigates the potential of Structural Similarity (SSIM) to bridge this gap. SSIM captures both intensity and structural disparities and can be advantageous over the classical $l1$ error. However, we show that there is more than one optimal kernel size for the SSIM calculation for different pathologies. Therefore, we investigate an adaptive ensembling strategy for various kernel sizes to offer a more pathology-agnostic scoring mechanism. We demonstrate that this ensembling strategy can enhance the performance of DMs and mitigate the sensitivity to different kernel sizes across varying pathologies, highlighting its promise for brain MRI anomaly detection.
\end{abstract}
\section{Introduction}
\label{sec:intro}

While supervised deep learning techniques have demonstrated encouraging outcomes in detecting and segmenting anomalies in brain MRI scans \cite{Lundervold.2019}, they depend on annotated and balanced data sets \cite{Johnson.2019}. The need for such large data sets, especially with voxel-level annotations, is a challenge given the labor and resource-intensive requirements. 
Unsupervised anomaly detection (UAD) methods take a different approach and only require healthy or normal data for training. In reconstruction-based UAD, a generative model (GM) is trained to replicate MRI scans from a healthy training set. Following this, discrepancies between an MRI scan and the GM's replication can be used to identify abnormal patterns in unhealthy brain scans. Hence, unlike supervised models, UAD techniques are not dependent on pixel-level annotations of diseases and have the invaluable potential to identify any kind of irregularity that differs from a norm learned from the healthy training distribution. \\
In the domain of UAD for brain MRI, various generative models (GMs) are used, with Autoencoders (AE) and their variational equivalents (VAE) being among the most common \cite{Baur.2021b}. 
Recently, denoising diffusion probabilistic models (DDPM) have emerged as promising options, offering precise reconstructions, effective representation of the training set, and stable training characteristics \cite{Wyatt.2022, Behrendt.2023,Behrendt.2023b}. 
% Problem: White object detectors
While the type and architecture of the GM play a crucial role in UAD, another vital design parameter is the type of discrepancy measurement used to score anomalies. Predominantly, methods rely on the mean squared error ($l2$-error) or mean absolute error ($l1$-error).  
%However, these metrics mainly offer intensity-based anomaly scoring. 
However, these metrics often miss structural differences and only focus on intensity-based discrepancies \cite{Meissen.2022,Meissen.2022b}. %, as illustrated in Figure \ref{fig:motivation}. 
Consequently, subtle anomalies of smaller intensity fluctuations might not be detected and reconstruction errors can be over-penalized.
\begin{figure}[t]
    \centering
    \includegraphics[width=\columnwidth]{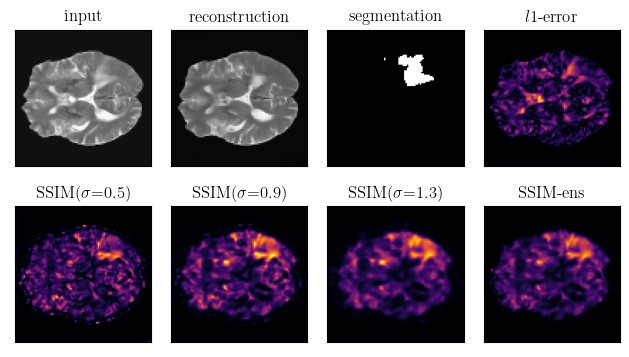}
    \caption{Visualization of $l1$-based and SSIM-based anomaly scores for different values of $\sigma$ and our ensemble solution SSIM-ens. Example is taken from the BraTS21 data set.}
    \label{fig:images}
\end{figure}
% SSIM for the rescue
An alternative anomaly score that enables the detection of both intensity-based and structural discrepancies is the Structural Similarity (SSIM) \cite{Wang.2004}. SSIM can provide a more balanced assessment, accounting for structural integrity, as demonstrated in Figure \ref{fig:images}.
Hence, existing literature suggests its application for UAD, either for industrial defect detection \cite{Bergmann.2018} or brain MRI pathology identification \cite{Meissen.2022b,Lagogiannis.2023}. Bergmann et al. \cite{Bergmann.2018} leverage a straightforward AE-based framework for detecting industrial defects, exchanging the $l2$-error with SSIM.  Meissen et al. \cite{Meissen.2022b} derive the anomaly score by computing the SSIM on reconstructed AE features rather than in the pixel space. \\
In this work, we investigate possible synergies of SSIM with the strong reconstruction abilities of recent DDPMs.

Our experimental results show that SSIM can improve the UAD performance when applied together with DDPMs. However, the kernel dimension $\sigma$ introduces an additional hyperparameter that limits the generalization across different pathology types. Our findings underscore that this parameter plays a pivotal role in the detection of pathologies, with distinct pathologies of certain sizes showing preferences for specific kernel dimensions.
To counteract this, we investigate utilizing an adaptive weighted average of multiple SSIM measurements over a spectrum of kernel sizes, mitigating the parameter sensitivity of SSIM. This method, which we call SSIM-ens, reduces the impact of individually chosen kernel sizes and leads to more robust performance in detecting a diverse range of pathologies, surpassing the results obtained with the traditional $l1$-error as anomaly score.

\section{Methods}
We adopt the reconstruction-based UAD methodology, selecting DDPMs as our GM of choice. The UAD principle involves training the DDPM to reconstruct MRI scans without anomalies. During testing, discrepancies between the input MRI scan and its reconstruction are flagged. These disparities often hint at structures not encountered during the training phase, suggesting potential abnormalities. 
\subsection{Diffusion Models}
DDPMs are generative models designed to capture the inherent data distribution of a given data set. In the training phase, DDPMs gradually transform an input image into Gaussian noise through a forward process, which is then reversed in the backward process to reconstruct the original image. This transformation is guided by a predefined schedule and involves adding controlled amounts of noise. The forward transformation is captured by:
% \[
$\bm{x}_t \sim q(\bm{x}_t|\bm{x}_0)$ %= \mathcal{N}(\sqrt{\bar\alpha_t} \bm{x}_0, (1-\bar\alpha_t) \textbf{I})$
% \]
where $t$ determines the noise level that is sampled from $t \sim Uniform(1,...,T)$ and $\bm{x}_0$ denotes a noise-free image. The backward process recovers the original image using:
% \[
$\bm{x}_{0}^{rec} \sim p(\bm{x}_T) \prod\nolimits^T_{t=1} p_{\theta}(\bm{x}_{t-1}|\bm{x}_t)$.
% \]
Similar to \cite{Behrendt.2023}, instead of predicting the added noise, we focus on directly estimating the reconstructed image, leading to the loss function:
% \[
$\mathcal{L}_{rec} = |\bm{x}_0 - \bm{x}^{rec}_{0}|$.
% \]
At test time, we directly estimate the reconstructed image based on the input, setting a fixed noise level $t_{test}$. 
\subsection{Anomaly Scoring with SSIM}
SSIM is designed to measure the local similarities of two images where high values of SSIM reflect similar images and low values indicate differences across the two images. Instead of solely calculating intensity-based discrepancies, SSIM incorporates changes in structural information, contrast, and luminance. The SSIM equation for two images \( \bm{x} \) and \( \bm{y} \) is given by
\[
SSIM(x,y) = \frac{(2\mu_x\mu_y + C_1)(2\sigma_{xy} + C_2)}{(\mu_x^2 + \mu_y^2 + C_1)(\sigma_x^2 + \sigma_y^2 + C_2)} 
\]
where \( \mu_x \) and \( \mu_y \) are the local means of \( x \) and \( y \), respectively; \( \sigma_x^2 \) and \( \sigma_y^2 \) are the local variances; and \( \sigma_{xy} \) is the local covariance. The constants \( C_1 \) and \( C_2 \) serve to stabilize the division with a weak denominator. The local statistics required for these computations are derived using a Gaussian kernel, with the \( \sigma \) parameter defining the kernel's spread. The kernel dimension is derived by multiplying the spread with a constant factor $k_{dim}=int(3.5 * \sigma + 0.5) * 2 + 1$. \\\\
\textbf{Adaptive Ensembling of SSIM (SSIM-ens)}\\
Calculating SSIM is sensitive to the spread of the Gaussian Kernel where larger values for $\sigma$ enlarge  the considered neighborhood for a given pixel and vice versa. This can pose a challenge, as it affects the scale at which discrepancies are detected.
To attenuate this dependency, we introduce SSIM-ens, an adaptive ensemble method that combines SSIM calculations over various \( \sigma \) values. This approach is intended to leverage the SSIM measurements at multiple scales, thereby providing an adaptive and more robust anomaly score.
For a given input image \( \bm{x}_{0} \) and its reconstruction \( \bm{x}_{0}^{rec} \), the SSIM-ens anomaly score is calculated as a weighted average across different \( \sigma \) values \( S = \{\sigma_1, \sigma_2, \ldots, \sigma_n\} \):
\[
\text{SSIM-ens}(\bm{x}_{0}, \bm{x}_{0}^{rec}) = 1 - \sum_{i=1}^{n} \bm{w}_i \cdot \text{SSIM}_{\sigma_i}(\bm{x}_{0}, \bm{x}_{0}^{rec}), \]
with 
\[
\bm{w}_i = \frac{e^{-\text{SSIM}_{\sigma_i}(\bm{x}_{0}, \bm{x}_{0}^{rec})}}{\sum_{j=1}^{n} e^{-\text{SSIM}_{\sigma_j}(\bm{x}_{0}, \bm{x}_{0}^{rec})}}
\]
where \( \text{SSIM}_{\sigma_i} \) is the individual SSIM calculated with the \( \sigma_i \) value, and \( \bm{w}_i \) is a normalized exponential weighting factor inversely related to the corresponding \( \text{SSIM}_{\sigma_i} \) score, enhancing the focus on areas with higher discrepancies and potential anomalies. This results in a scoring mechanism that effectively captures variations in anomaly manifestation across different pathology types and sizes.
\section{Experimental Setup}
% Data
\subsection{Data Sets}
We utilize the Information eXtraction from Images (IXI) data set that contains healthy brain MRI scans as the training set. The IXI data set includes 560 pairs of T1- and T2-weighted MRI scans. We sample a healthy test set (N$_{test}$=158) and partition the remaining samples into five healthy training sets (N$_{train}$=358) and 5 healthy validation sets (N$_{val}$=44) for five-fold cross-validation.\\
For evaluation, we rely on pathology data sets that provide pixel-wise annotations of pathology regions. Namely, we utilize the BraTS21, the ATLAS (v2), the WMH and MSLUB data sets, containing tumors, stroke, white matter hyperintensities and multiple sclerosis lesions, respectively. For the BraTS21 and MSLUB data sets, 1251 and 30 T2-weighted MRI scans and for the ATLAS and WMH data sets, 655 and 60 T1-weighted samples are available, respectively. For each evaluation data set, we split an unhealthy validation set of 100, 10, 175 and 15 samples for BraTS21, MSLUB, ATLAS and WMH respectively and use the remaining data as test set. \\
Pre-processing includes resampling to an isotropic resolution of $1\mathrm{\,mm}\times 1\mathrm{\,mm} \times 1\mathrm{\,mm}$ and affine registration to the SRI24 Atlas. Subsequently, we skull-strip the brain scans and perform N4 bias field correction. Lastly, the volume resolution is reduced by a factor of two and the 15 top and bottom slices parallel to the transverse plane are removed, leading to a final resolution of $96 \times 96 \times 50$\, voxels. 
For post-processing of the anomaly maps, median filtering with a kernel size of 5x5x5, brain mask eroding and connected components analysis is applied, similar to \cite{Baur.2021b,Behrendt.2023}. For binarization, we calculate a threshold that shows the highest segmentation performance based on the unhealthy validation sets.  

\subsection{Implementation Details}
We utilize a Unet, with channel dimensions of [128, 256, 256] as a denoising network\footnote{Code available at \url{https://github.com/FinnBehrendt/Ensembled-SSIM-for-Unsupervised-Anomaly-Detection}}. We utilize structured simplex noise, which enhances the UAD efficacy of DDPMs in MRI images \cite{Wyatt.2022}. For the sampling process, we sample \( t \) uniformly from the interval \([0, 999]\) during training. At test time, we employ a consistent value for \( t \) by setting \( t_{test} = 500 \). For the SSIM-ens, we utilize a range of \( \sigma \) values \( S = \{0.3, 0.5, ..., 1.5, 1.7\} \).
We utilize the Adam optimization algorithm, with a learning rate of $1 \times 10^{-5}$ and a batch size of 32. After training for 1600 epochs, model selection is based on the lowest reconstruction error achieved on the healthy validation set. Data processing is performed slice-by-slice; during training, slices are randomly selected with replacement, while at test time, we iterate through every slice to reconstruct the entire volume. 
In addition to the DDPMs, we implement AEs \cite{Baur.2021b} and denoising AEs (DAE) \cite{Kascenas.2022b} as baselines.
\section{Results}
\begin{figure}
    \centering
    \includegraphics[width=\columnwidth]{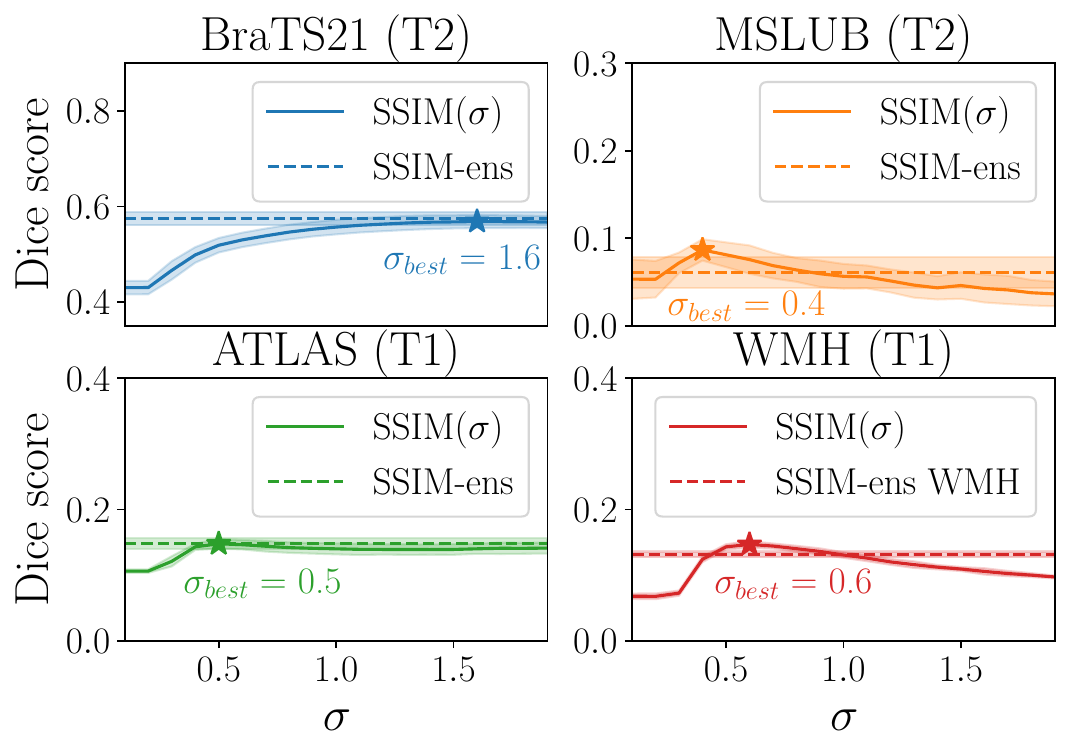}
\caption{Assessment of segmentation performance of DDPMs with SSIM using the Dice coefficient across varying SSIM parameter \(\sigma\) values. Optimal performance instances are denoted by stars ($\sigma_{best}$). The solid lines delineate the mean performance, while the surrounding shaded regions depict the standard deviation. The performance achieved by our SSIM-ens method is illustrated with dashed lines.}
    \label{fig:Dice_vs_sigma}
\end{figure}
\begin{table}[h!]
\centering
\caption{Comparison of the evaluated models with color-coded changes within each model group. For all metrics, the mean $\pm$ standard deviation across the different folds are reported.  }
\resizebox{\linewidth}{!}{
\begin{tabular}{lcccccccccc}
\toprule
 &  \multicolumn{1}{c}{BraTS21 (T2)}  & \multicolumn{1}{c}{MSLUB (T2)} & \multicolumn{1}{c}{ATLAS (T1)} & \multicolumn{1}{c}{WMH (T1)} \\ 
\textbf{Model}  & \textbf{DICE [\%]} &  \textbf{DICE [\%]} &  \textbf{DICE [\%]} &  \textbf{DICE [\%]} \\
\midrule
\textit{AE} \cite{Baur.2021b} ($l1$) & 31.51$\pm$1.94  &   7.23$\pm$0.90 &   14.91$\pm$0.33 &  4.53$\pm$0.36 \\
\textit{AE} \cite{Baur.2021b} (SSIM-ens) & \textcolor{red}{$\downarrow$ 25.88$\pm$0.43  } & \textcolor{red}{$\downarrow$ 3.57$\pm$0.70}  &   \textcolor{red}{$\downarrow$ 8.23$\pm$0.13 } & \textcolor{darkgreen}{$\uparrow$ 6.53$\pm$0.20 } \\ \midrule
\textit{DAE} \cite{Kascenas.2022b} ($l1$) & 45.37$\pm$4.40 &  3.88$\pm$1.35 &   8.53$\pm$0.28 &  7.31$\pm$1.02 \\
\textit{DAE} \cite{Kascenas.2022b} (SSIM-ens) & \textcolor{darkgreen}{$\uparrow$ 59.24$\pm$0.63 }  & \textcolor{red}{$\downarrow$ 1.83$\pm$0.16 }  &  \textcolor{darkgreen}{$\uparrow$ 15.03$\pm$0.52 }   & \textcolor{darkgreen}{$\uparrow$ 8.76$\pm$0.38} \\\midrule
\textit{DDPM} \cite{Wyatt.2022} ($l1$) & 44.25$\pm$1.49 &   4.80$\pm$1.98 &   12.90$\pm$0.89  & 10.03$\pm$1.06 \\
\textit{DDPM} \cite{Wyatt.2022} (SSIM-ens) & \textcolor{darkgreen}{$\uparrow$ 57.44$\pm$1.40 }&  \textcolor{darkgreen}{$\uparrow$ 6.10$\pm$1.78  }& \textcolor{darkgreen}{$\uparrow$ 14.81$\pm$0.79 }  &  \textcolor{darkgreen}{$\uparrow$ 13.16$\pm$0.44}\\
\bottomrule
\end{tabular}
}
\label{tab:results}
\end{table}
\noindent Initially, we explore the impact of the \(\sigma\) parameter of SSIM on the segmentation performance of DDPMs for various pathologies. Following this, we evaluate and compare baseline models, both with and without the integration of our SSIM-ens strategy. To evaluate the segmentation performance, we report the pixel-wise Dice coefficient (DICE).\\
In Figure \ref{fig:Dice_vs_sigma} we observe that the \(\sigma\) parameter affects the segmentation performance across all data sets. Selecting an optimal singular \(\sigma\) value is not straightforward due to the presence of multiple optimal points that vary between data sets. In contrast, by implementing an adaptive ensemble of diverse \(\sigma\) values within SSIM-ens, consistent performance is maintained across all data sets.\\
In Table \ref{tab:results}, it is evident that applying the SSIM-based anomaly score leads to notable performance improvements for DDPMs across all data sets, reliably surpassing the $l1$-error. 
\section{Discussion and Conclusion}
In this work, we provide an evaluation of the SSIM as an anomaly score for UAD in brain MRI. We demonstrate that the performance of SSIM is tied to the values of the kernel dimensions specified by $\sigma$, where the optimal values of $\sigma$ vary for individual pathology types and sizes. Our study investigates an extension, SSIM-ens, which mitigates the parameter-dependence problem and offers a robust solution when integrated into DDPMs.\\
We show that our ensemble strategy consistently outperforms traditional \(l1\) anomaly scores across varying pathologies if applied to DDPMs. By taking a weighted average of different SSIM anomaly scores across multiple parameter settings, SSIM-ens adds a degree of universality to the SSIM since the weighting is determined by the SSIM scores themselves, which means that the method adapts to the data it's evaluating. For discrepancies that are more prominent at certain scales, those discrepancies will be given more weight in the final score.  \\
A notable observation is that SSIM does not consistently improve the performance of AEs and DAEs. For AEs the reason is seen in blurry reconstructions, failing to accurately mimic normal anatomy. On the other hand, DAEs, due to their inherent biases, often reconstruct pathologies different from tumor-like anomalies, as indicated in \cite{Behrendt.2023}. This impacts the effectiveness of SSIM-ens, which depends on precise representations of healthy anatomy in both the input and the reconstructed output.\\
Overall, our findings underscore the significance of the choice of anomaly score metric. It appears that the type of anomaly score utilized can have a substantial impact on UAD performance, potentially overshadowing the effects of changing the architecture of the generative model. 
We show that even though increasing performance across different data sets, SSIM alone adds a parameter dependence that prevents an optimal solution for differing pathologies. The proposed ensembling strategy can reduce the pathology-specific search for hyper-parameters and offers a more general solution for UAD in brain MRI.
\\\\
\noindent\textbf{Ethical approval:} This research study was conducted using public data. Therefore, ethical approval was not required.
\\\\
\textbf{Funding:} This work was partially funded by grant number KK5208101KS0 and  by the Free and Hanseatic City of Hamburg (Interdisciplinary Graduate School). %

\label{sec:acknowledgments}

\bibliographystyle{IEEEbib}
\bibliography{refs}

\end{document}